\documentclass{aastex62}

\usepackage{units}
\usepackage{mathtools}
\usepackage{bm}

\graphicspath{{./}{figures/}}

\received{May 20, 2019}
\revised{June 20, 2019}
\accepted{June 22, 2019}
\submitjournal{ApJ}

\shorttitle{TESS Search for FRB181228}
\shortauthors{Tingay \& Yang}

\begin{document}

\title{A Search of TESS Full Frame Images for a Simultaneous Optical Counterpart to FRB181228}

\correspondingauthor{Steven Tingay}
\email{s.tingay@curtin.edu.au}

\author[0000-0002-8195-7562]{Steven J Tingay}
\affil{International Centre for Radio Astronomy Research, Curtin University, Bentley, WA, Australia 6102; s.tingay@curtin.edu.au}

\author{Yuan-Pei Yang}
\affil{South-Western Institute for Astronomy Research, Yunnan University, Kunming, Yunnan, P.R. China; ypyang@ynu.edu.cn}



\begin{abstract}
FRB181228 was detected by the Molonglo Synthesis Radio Telescope (MOST) at a position and time coincident with Transiting Exoplanet Survey Satellite (TESS) observations, representing the first simultaneous multi-wavelength data collection for a Fast Radio Burst (FRB).  The large imaged field-of-view of TESS allows a search over the uncertainty region produced by MOST.  However, the TESS pixel scale of 21\arcsec~ and the Full Frame Image (FFI) cadence of 30 minutes is not optimal for the detection of an FOB with a possible millisecond duration.  We search the TESS FFIs and find no events with a limiting TESS magnitude of 16, assuming a 30 minute event duration, corresponding to an optical flux density upper limit of approximately 2000$\,$Jy for a $\sim 1\,$ms signal duration, assuming no signal loss.  In addition, the cosmic ray mitigation method for TESS significantly reduces its sensitivity to short timescale transients, which we quantify.  We compare our results to the predictions of \citet{yan19} and find that the upper limit is a factor of two thousand higher than the predicted maximum optical flux density.  However, we find that if FRB181228 had occurred in the galaxy thought to host the nearest FRB detection to date (37 Mpc), an FOB may have been detectable by TESS.  In the near future, when CHIME and ASKAP will detect hundreds to thousands of FRBs, TESS may be able to detect FOBs from those rare bright and nearby FRBs within this large population (if more sophisticated cosmic ray excision can be implemented).
\end{abstract}



\section{Introduction} \label{sec:intro}

Fast Radio Bursts (FRBs) are short duration ($\sim$millisecond) and intense bursts of radio emission, originating from cosmological distances as inferred from their high dispersion measures.  The first FRB was detected in archival data from the Parkes Radio Telescope by \cite{lor07}.  Since then, a range of radio telescopes have detected FRBs.

The first robust identification of an FRB host galaxy came from observations of a repeating FRB, FRB121102, the repeating nature of the object greatly assisting with the localisation and host galaxy identification \citep{cha17,mar17,bas17,ted17}, placing the FRB some 3 billion light years distant.  The Australian SKA Pathfinder (ASKAP) telescope is detecting FRBs at a high rate \cite{askap}.  While ASKAP has localisation capability, no localisations have yet been published.  

The recently operational Canadian Hydrogen Intensity Mapping Experiment (CHIME) telescope \citep{ban14,chimefrb} has already formally reported thirteen FRB detections over an observing period of less than one month \citep{chimefrb19}.  CHIME is not currently capable of localising FRBs to a level that host galaxies can be easily identified, although upgrades to enable this capability are in progress.  The refurbished Molonglo Observatory Synthesis Telescope (MOST), utilised for the UTMOST project, has also detected a number of FRBs, most recently reported by \citet{utfrb}, as has the Parkes Radio Telescope, most recently reported by \citet{parkes}.  The Green Bank Telescope (GBT) has reported one FRB \citep{gbt} and Arecibo has reported two FRBs \citep{arecibo1,arecibo2}.  In all these cases, localisation precision is low.  Radio telescopes working at low radio frequencies ($<$300 MHz) are yet to detect any FRBs, despite some efforts to shadow telescopes operating at higher frequencies while the later achieve detections \citep{sok18}.  A useful database to track the current state of published FRB detections is available online at http://frbcat.org \citep{pet17}.

The observational characteristics of FRBs are complex and are currently not comprehensively understood.  One of the fundamental observational challenges to obtaining further physical insight is their appearance at random times and positions on the sky.  Thus, while wide-field telescopes such as CHIME and ASKAP may detect and even localise FRBs in real-time, obtaining simultaneous multi-wavelength information on an FRB is extremely difficult.  Very wide field-of-view optical telescopes are required to constantly shadow the radio telescopes, imaging at high cadence, to obtain simultaneous sky coverage with appropriate temporal resolution.  In the vast majority of cases, this is not practical.

However, for very wide field-of-view optical telescopes tasked with other missions (but not optimised for transients), it is possible that by chance they concurrently cover the part of sky in which a radio telescope detects an FRB.  This is the case reported here where, for the first time, optical data have been collected simultaneously with radio observations of an FRB, using the Transiting Exoplanet Survey Satellite (TESS; \citet{tess}).  TESS is designed for precision photometry aimed at the detection of exoplanet transits toward relatively bright stars.

The TESS observing field-of-view covers $24^{\circ} \times 96^{\circ}$, spanning from the ecliptic equator to an ecliptic pole (6$^{\circ}$ to 96$^{\circ}$ in ecliptic latitude), with a dwell time of approximately 27 days per each of 26 Sectors (thirteen in the ecliptic Southern Hemisphere and thirteen in the ecliptic Northern Hemisphere).  In year one, the Southern Hemisphere was covered, with year two covering the north.   For a catalogue of approximately 20,000 targets, TESS produces image cutouts with a 2 minute cadence ($<$0.01\% of the sky at any given time), for variability studies.  However, each Sector is continuously covered by Full Frame Images (FFIs) with a 30 minute cadence.  All TESS images have a pixel scale of 21\arcsec~ and 90\% of light from a point source is contained in a 2$\times$2 pixel area.  Full details of the TESS observing strategy and instrument characteristics can be found in the TESS Instrument Handbook\footnote{https://archive.stsci.edu/files/live/sites/mast/files/home/missions-and-data/active-missions/tess/\_documents/TESS\_Instrument\_Handbook\_v0.1.pdf
}.

Although the ultra-wide field of the TESS observing system affords a unique opportunity for transient studies, the 21\arcsec~ pixel size and 30 minute cadence of the FFIs represent very significant limitations, for FRBs in particular.  Assuming that the optical emission from an FRB has a similar timescale to the radio emission ($\sim$ms), a 30 minute exposure represents a temporal dilution of the signal by a factor of approximately 1.8 million, meaning only very bright short duration signals can conceivably be detected.

However, within the TESS observations of Sector 6, an FRB was detected by the UTMOST project, FRB181228 \citep{utfrb}.  The TESS data coincident with the UTMOST detection represent the first simultaneous optical and radio observations of an FRB.  With the limitations mentioned above in mind, this represents a unique situation, worthy of report.  Moreover, as TESS has largely covered only southern declinations in year one, the result reported here is likely to be replicated many times over once TESS covers largely northern declinations in year two, given the high detection rate by the CHIME telescope.  With the large field-of-view of CHIME, the rare nearby and/or extremely bright FRBs detected may have a significantly increased chance of detection with the TESS system.  Thus, one purpose of this paper is to highlight future near-term possibilities for TESS in FRB research.

A multitude of theories and models exist to explain the phyiscal nature of FRBs.  The reader is referred to \cite{platt19} for an extensive and general theory catalogue for FRBs.  In this paper, we restrict ourselves to a discussion of FRB theories of relevance to optical emission.  In particular, we reference the recent predictions for optical emission accompanying FRBs, so-called Fast Optical Bursts (FOBs), by \citet{yan19}.  These authors investigate theoretical predictions for a range of physical scenarios for optical emission from FRBs, covering variants of Inverse Compton (IC) processes and processes that extend the proposed radio emission mechanisms into the optical band.  For a fiducial 1$\,$Jy FRB, they find upper limits of $\sim$0.01$\,$Jy in the optical band for the associated FOB, and compare these limits to future wide field, high imaging cadence optical telescope capabilities, primarily for LSST.

In \S 2, the TESS data processing for FRB181228 is described, along with the results of the processing in the form of upper limits on the optical emission from FRB181228.  In \S 3 we discuss the relevance of the current results for models of FRBs, via a comparison to the \citet{yan19} predictions for the optical emission from FRB181228.  We also discuss future prospects for FRB research with TESS (and potentially other similar missions).

\section{Data Processing and Results} \label{sec:data}

FRB181228 occured at UTC 2018-12-28-13:48:50.1 and was detected as part of the UTMOST program at MOST between 820 and 850$\,$MHz \citep{utfrb}.  The FRB had a DM of 354.2$\,$pc$\,$cm$^{-3}$, placing the FRB at a maximum redshift of 0.3 after the galactic contribution to the DM was estimated and removed.  The fluence was $>$24$\,$Jy ms and the duration of the FRB was 1.24$\,$ms.  The peak flux density was approximately 19$\,$Jy.

Due to the characteristics of MOST, the localisation error region is highly elongated in a north-south direction, with an east-west extent of only 5\arcsec~ but a north-south extent of several degrees (1$\sigma$ of 1.2$^{\circ}$).  The most accurate position for FRB181228 is RA: 06:09:23.64; Dec: $-$45:58:02.4 (J2000), and the localisation arc has the following equation:

\[\alpha = 6.156567 + 1.057078\times10^{-3} (\delta + 45.967333) - 2.784623\times10^{-5} (\delta + 45.967333)^{2}\]

where $\delta$ is in the range $-$50 to $-42$] $\alpha$ is in decimal hours of RA, and $\delta$ is in decimal degrees of Dec \citep{utfrb}.

The position and time of FRB181228 occured within TESS Sector 6 observations.  A python script based on the extraction of light curves from TESS cutouts of calibrated FFIs\footnote{https://spacetelescope.github.io/notebooks/notebooks/MAST/TESS/interm\_tesscut\_astroquery/interm\_tesscut\_astroquery.html} was developed.  The script loops over the range $-50^{\circ} < \delta < -42^{\circ}$ and evaluates the above equation to generate a series of $\alpha$ and $\delta$ positions at which to produce FFI cutouts of 10 $\times$ 10 pixels.  These cutouts were downloaded by the script for further processing.

The flux (in units of electrons per second) in each cutout was integrated across all 100 pixels and was corrected for background variations by examining the dimmest pixels (those in the bottom 5\% in flux) and subtracting this background after accounting for the number of pixels utilised in forming the background measurement.  From these background corrected integrated fluxes, light curves were formed at each position (1372 positions covering the localisation arc over $-50^{\circ} < \delta < -42^{\circ}$).  The light curves were restricted to contain points within $\pm0.2$ days of the FRB time, comprising 19$\times$30 minute FFIs.  An example light curve is shown in Figure 1, corresponding to the most accurate position for FRB181228.  All light curves at all positions along the localisation error region appear qualitatively similar, thus we only present this one light curve as an example, not because there is an overwhelmingly greater chance of finding a signal at the best fit FRB position.

\begin{figure}[ht!]
\plotone{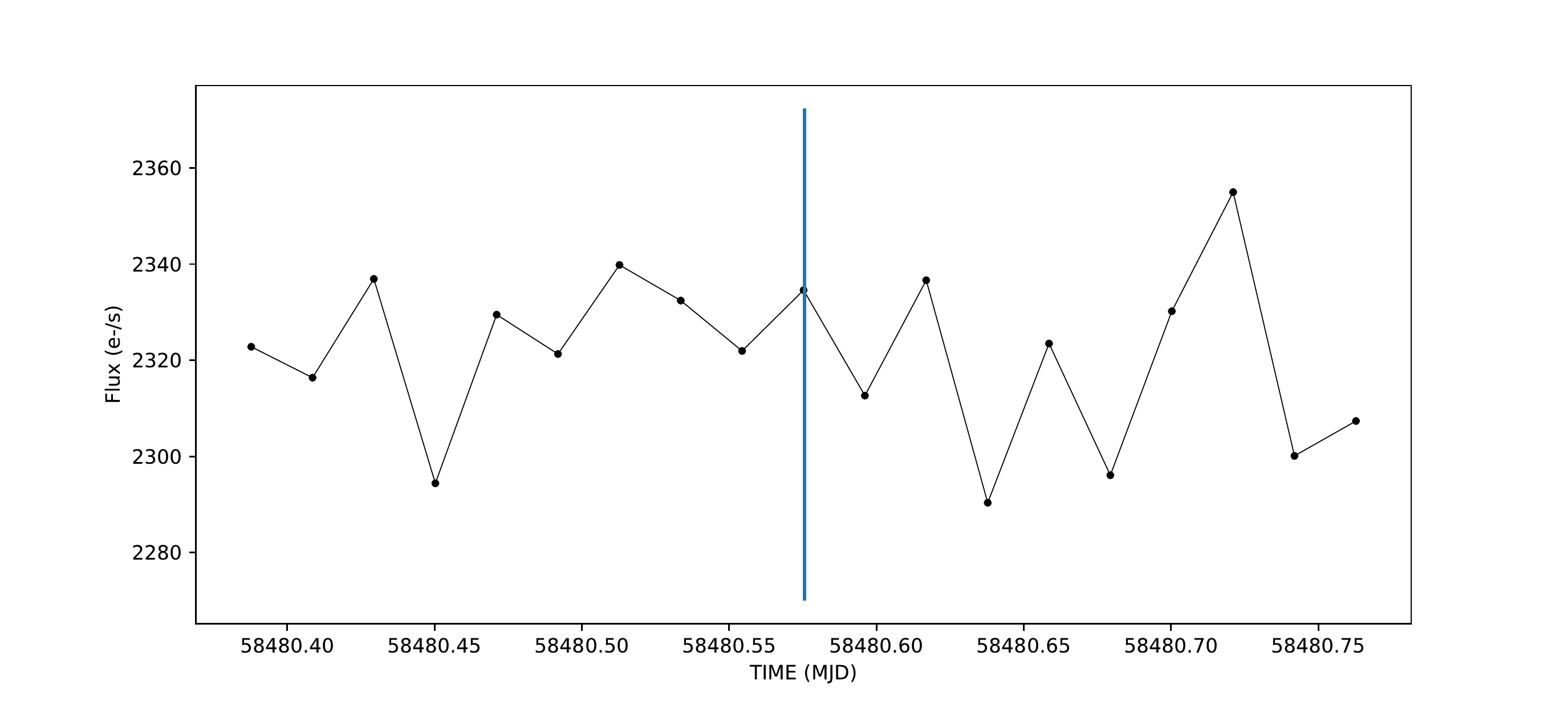}
\caption{An example TESS light curve at the best fit position for FRB181228 from \cite{utfrb}.  The vertical blue line denotes the time of the FRB and the vertical extent is $\pm3$ times the RMS around the mean flux.\label{fig:1}}
\end{figure}

For each light curve, the mean and RMS were calculated, with the RMS at most only a few percent of the mean in all cases, indicating highly stable flux measurements.  This was confirmed with visual inspection of the light curves.  Each light curve was evaluated to determine if a variation from the mean greater than three times the RMS occurred at the time of the FRB.  No such event was seen across the 1372 light curves.  Data points more than three times the RMS from the mean were seen in 6/1372 light curves, at times not corresponding to the FRB time, at: $-5\,$hr ($\times$2); $-4.5\,$hr; $-2\,$hr; and $+1.5\,$hr ($\times$2, but due to the same data appearing in light curves from two adjacent positions on the sky).  This rate of events above three sigma from the mean is broadly consistent with the expectation from Gaussian distributed data for the number of points per light curve and the number of light curves, given that the light curves are not fully independent (the centroid of the $10\times10$ pixels advances one pixel in declination between light curves).  Alternatively, some of these events could represent low level cosmic ray affected pixels that are not removed by the cosmic ray mitigation process for TESS FFIs.  

The cosmic ray mitigation process for TESS complicates the detection of short duration transient events.  The highest and lowest values of each pixel from groups of 10$\times$2 second images are excised before integration into a 20 second image (with the resulting sensitivity of a 16 second image).  For the FFIs utilised here, these 20 second images are then integrated into 30 minute images (with the sensitivity of a 24 minute image).

This method of cosmic ray mitigation means that if an astrophysical signal of less than 2 seconds in duration is present in a 20 second period and produces the brightest pixel value for that pixel during that period, it will be removed by the mitigation system.  The chance of a pixel experiencing a cosmic ray during a 20 second period is approximately 1.7\% \citep{lun16}.  If a cosmic ray produces the brightest pixel in a 20 second period, short term transients at lower fluxes may be preserved 1.7\% of the time.  This represents a very limited and caveated TESS sensitivity to transients of $<$2 seconds duration.  

However, if the transient duration is $>$2 seconds, the majority of the signal is preserved most of the time, since only one 2 second pixel can be excised every 20 seconds.  Figure 2 shows the maximum signal loss as a percentage, for transient durations up to 60 seconds.  Note that the maximum signal loss of 100\% for 4 second duration transients is due to the scenario in which both the last 2 second pixel of a 20 second group and the first 2 second pixel of the next 20 second group are excised, which may be the case for up to 10\% of 4 second transients.  If the adjacent 2 second pixels exist within the same 20 second period, the maximum signal loss is 50\%, which will be the case for 90\% of 4 second transients.  Similar effects exist at lower levels for other transient durations.

\begin{figure}[ht!]
\plotone{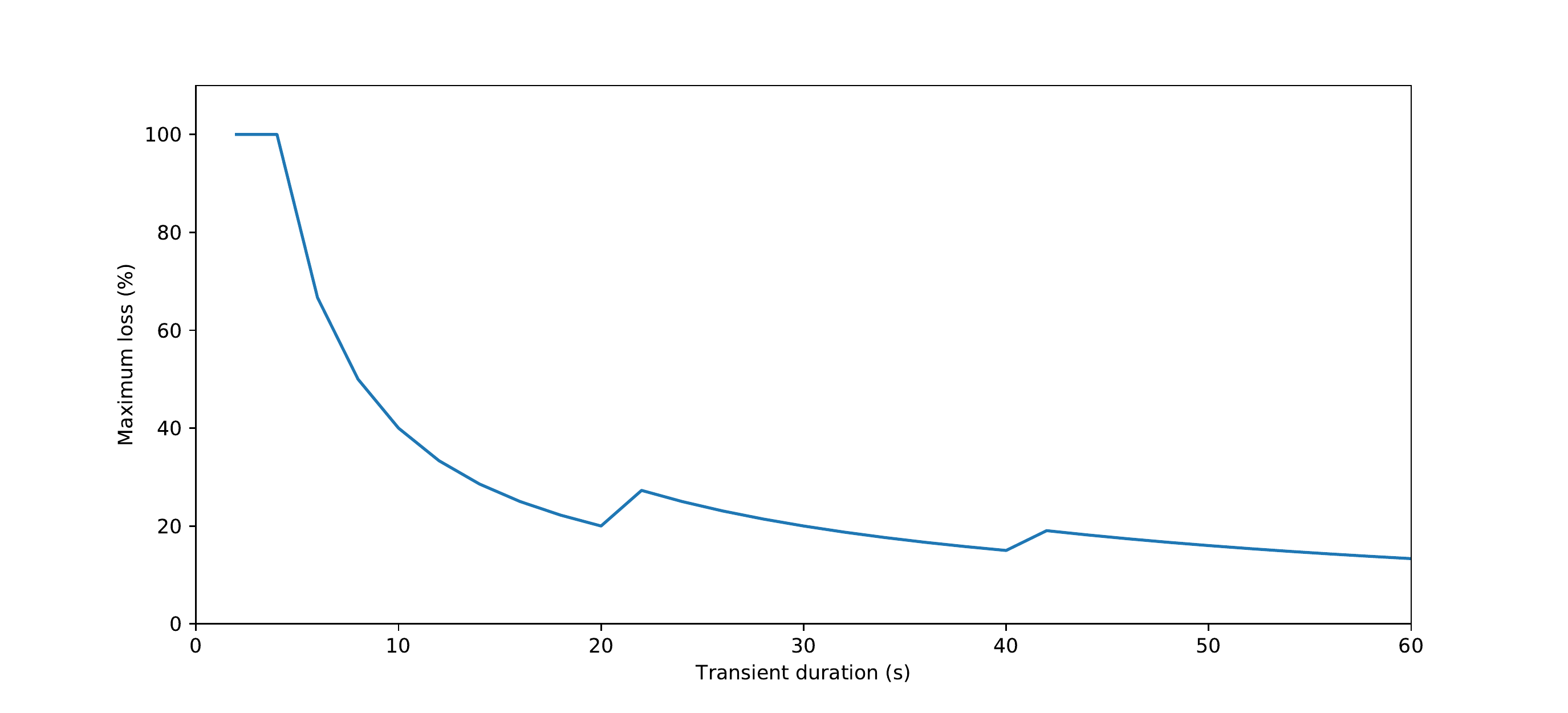}
\caption{\label{fig:2} {\bf Maximum signal loss as a function of transient durations up to 60 seconds, due to the TESS cosmic ray mitigation process.}}
\end{figure}

The average RMS value for the flux in our light curves was approximately 15$\,$e$^{-}$/s, thus a three sigma value of 45$\,$e$^{-}$/s above the mean represents close to our best (and typical) upper limit, given that we found no signal exceeding the three sigma threshold at the FRB time in any light curve.  From the TESS Instrument Handbook, a flux of 15,000$\,$e$^{-}$/s corresponds to a TESS magnitude of 10.  Thus, a limit on excess flux at the FRB time of 45$\,$e$^{-}$/s corresponds to no more than a 16$^{\rm th}$ magnitude object appearing for the duration of the relevant 30 minute FFI.

As noted in the introduction, the 30 minute FFI cadence represents a limitation for FRB searches with TESS, since the upper limit of 16$^{\rm th}$ magnitude applies to a constant flux over that time period.  Assuming that the optical emission duration matches the 1.24$\,$ms radio emission duration, the upper limit for this much shorter duration is raised to a TESS magnitude of approximately 0.5.  However, as noted above, a $\sim$1 ms event has only a very small chance of being detectable, on cosmic ray mitigation considerations.  If the optical emission has a longer duration than the radio emission, the limits are reduced.  For example, for the shortest plausible detection duration of 4 seconds, the TESS magnitude upper limit for a transient of this duration is approximately 10, taking into account the most likely 50\% signal loss.

\section{Discussion and Conclusions} \label{sec:disc}

As discussed above, the cosmic ray mitigation process for TESS makes detection of very short timescale transients difficult.  It is possible that a more subtle method of excising cosmic rays from TESS pixels could improve this situation.  For example, due to the TESS pixels being large and deep, cosmic rays can routinely affect significant numbers of adjacent pixels, rather than a single pixel \citep{lun16}.  If filters are designed to recognise adjacent pixels affected by cosmic rays, a larger fraction of short timescale transients could be recoverable. Therefore, if the sensitivity to transients of less than 2 seconds can be recovered in the TESS system, one would have a chance to detect/constrain FOB emission with duration less than 2 seconds.  The TESS instrument team has this possibility under consideration (\S9.2.2, TESS Instrument Manual), which the following analysis shows would be worthwhile.

We recognize that any optical emission mechanisms associated with FRBs are highly uncertain. Coupled with the fact that observational limits derived from TESS are unlikely to strictly constrain theoretical models, we therefore do not undertake an exhaustive comparison to multiple theoretical models, but restrict ourselves to one set of illustrative models elaborated by \citet{yan19}.

According to our analysis, the magnitude of the FOB associated with FRB 181228 is $m>16~{\rm mag}$ for the exposure time $T=30~{\rm min}$. 
Thus, if the sensitivity to transients with duration less than 2 seconds can be recovered in the TESS system, the optical flux density is given by \citet{yan19}:
\begin{eqnarray}
F_{\nu}=
\left(\frac{T_{60}}{\tau_{\rm ms}}\right)10^{(8.32-0.4m)}~{\rm Jy}<2012~{\rm Jy}\left(\frac{\tau_{\rm ms}}{1.24 {\rm ms}}\right)^{-1}\label{fobflux},
\end{eqnarray}
for the case in which the optical pulse duration satisfies $\tau\lesssim T$, where $\tau_{\rm ms}$ is the optical pulse duration in milliseconds and $T_{60}$ is the exposure time normalized to $60~\unit{s}$. Thus, if the FOB duration is the same as that for FRB 181228, the optical flux density upper limit is $F_\nu<2012~{\rm Jy}$. If the FOB duration reaches $\sim10^3~{\rm s}$ (but less than 30 min), the optical flux density is $F_\nu\lesssim2\times10^{-3}~{\rm Jy}$. However, if the FOB duration is larger than the exposure time, the optical flux density would not depend on the FOB duration, and one has $F_\nu=3631~{\rm Jy}\times10^{-0.4m}<10^{-3}~{\rm Jy}$ for $m>16~{\rm mag}$. 

An FOB could be produced by the IC scattering process or by the same mechanism that produces the FRB radio emission \citep{yan19}. Both for a one-zone IC model and for the FRB radio emission mechanism model, the predicted FOB duration is the same as that for the FRB. But for a two-zone IC scattering process, the predicted FOB duration could be longer, due to an extended IC scattering region. 

First, we consider that an FOB has the same duration as an FRB, and the sensitivity to transients with duration less than 2 seconds is assumed to have been recovered in the TESS system.
Based on the models explored in \citet{yan19}, for FRB 181228 with the radio flux density of $\sim19~{\rm Jy}$, we predict that the optical flux density is $\lesssim1~{\rm Jy}$, with the highest predicted optical flux density from the one-zone IC scattering in a pulsar magnetosphere model. 
The highest predicted optical flux is approximately a factor of two thousand below the upper limit we estimate from TESS data, $F_\nu<2012~{\rm Jy}$.

In the era of ASKAP and CHIME, through detecting and localizing hundreds or thousands of FRBs in short periods of time, one might expect that far brighter and/or nearby FRBs could be detected.  Potentially the closest FRB yet detected is the lowest DM detection by ASKAP, FRB171020, most likely identified with the galaxy ESO 601$-$G036 at a redshift of 0.00867 ($\sim$37 Mpc) \citep{mah17}.
If FRB181228 occured in ESO 601$-$G36, its radio flux density would be approximately 1900 times higher, or approximately 35.5 kJy. In this situation, the predictions of \citet{yan19} for the associated maximum optical emission can reach 1900 Jy, which would be very close to the TESS limit.  
Furthermore, FRB 180714 is one of FRBs with the highest luminosity. Its estimated maximum redshift is 1.35 ($\sim9.7$ Gpc) and its peak flux density is $5$ Jy \citep{osl18,zha18}. If FRB 180714 occured in ESO 601$-$G36, its radio flux density would be $344$ kJy, and the associated maximum optical emission can reach 18 kJy, which would be significantly larger than the TESS limit.
Thus, even with the limits imposed by the TESS pixel size and FFI imaging cadence, FOB detection for some rare bright and nearby FRBs may be possible, if sensitivity to transients of less than 2 seconds can be recovered in the TESS system.


Second, an FOB could be produced in a two-zone IC scattering process, which would cause a duration longer than that of the associated FRB due to the larger extent of the IC emission region.
In particular, for an FOB with duration $\gtrsim2$ s, the cosmic ray mitigation effect for TESS would not be significant, and the intrinsic optical emission can be directly obtained by Eq.(\ref{fobflux}).
For example, if an FOB is produced by a two-zone IC scattering process in a young supernova remnant (SNR) associated a pulsar, the predicted FOB duration could reach $\sim10^4~{\rm second}$ for a young SNR with an age of a few years \citep{yan19}. In this case, due to the predicted FOB duration being longer than the TESS exposure time, the optical flux density would be $F_\nu=3631~{\rm Jy}\times10^{-0.4m}<10^{-3}~{\rm Jy}$ for $m>16~{\rm mag}$. The optical flux density constrained by TESS in this case is lower than the predicted maximum possible optical flux density, $F_\nu\lesssim0.17~{\rm Jy}$, in the pulsar nebula (e.g. SNR) model.

The data and analyses presented here show that, while TESS is far from optimised for the detection of optical bursts associated with FRBs, in an era when many FRBs are being detected, it is possible that TESS may detect a small number of very bright and nearby FOBs associated with FRBs from FFIs.  

Future FRB detection experiments could easily target TESS observing regions, to ensure maximum overlap between radio and optical coverages.

\acknowledgments
This research has made use of NASA's Astrophysics Data System.  The authors thank the referee, Prof. Shri Kulkarni, for offering comments that significantly improved this manuscript.

%

\vspace{5mm}
\facilities{TESS}






\end{document}